# Electron Precipitation Observed by ELFIN Using Proton Precipitation as a Proxy for Electromagnetic Ion Cyclotron (EMIC) Waves

**L. Capannolo[1], W. Li[1], Q. Ma[1,2], M. Qin[1], X.-C. Shen[1], V. Angelopoulos[3], A. Artemyev[3], X.-J. Zhang[3,4], and M. Hanzelka[1,5]**

[1]Center for Space Physics, Boston University, Boston, MA.

[2]Department of Atmospheric and Oceanic Sciences, University of California, Los Angeles, CA.

[3]Department of Earth, Planetary, and Space Sciences, University of California, Los Angeles, CA.

[4]Department of Physics, The University of Texas at Dallas, Richardson, TX.

[5]Department of Space Physics, Institute of Atmospheric Physics of the Czech Academy of Sciences, Prague, Czech Republic

Corresponding authors: Luisa Capannolo (luisacap@bu.edu), Wen Li (wenli77@bu.edu)

## Key Points

- Energetic electron precipitation is observed by ELFIN nearby proton precipitation (a proxy for EMIC waves) primarily over 15–24 MLT

- Precipitation efficiency increases as a function of energy: weak ~100s keV precipitation is concurrent with intense ~MeV precipitation

- The observed pitch-angle distribution shows a loss cone filling up with energy, similar to the pitch-angle profiles from quasilinear theory

## Abstract

Electromagnetic Ion Cyclotron (EMIC) waves can drive radiation belt depletion and Low-Earth Orbit (LEO) satellites can detect the resulting electron and proton precipitation. The ELFIN (Electron Losses and Fields InvestigatioN) CubeSats provide an excellent opportunity to study the properties of EMIC-driven electron precipitation with much higher energy and pitch-angle resolution than previously allowed. We collect EMIC-driven electron precipitation events from ELFIN observations and use POES (Polar Orbiting Environmental Satellites) to search for 10s–100s keV proton precipitation nearby as a proxy of EMIC wave activity. Electron precipitation mainly occurs on localized radial scales (~0.3 L), over 15–24 MLT and 5–8 L shells, stronger at ~MeV energies and weaker down to ~100–200 keV. Additionally, the observed loss cone pitch-angle distribution agrees with quasilinear predictions at ≳250 keV (more filled loss cone with increasing energy), while additional mechanisms are needed to explain the observed low-energy precipitation.

## Plain Language Summary

Electromagnetic Ion Cyclotron (EMIC) emissions are a type of plasma wave that can be excited in the near-Earth environment and interact with energetic electrons in the Earth's radiation belts. Through these wave-particle interactions, electrons can be pushed into the loss cone and lost into the Earth's atmosphere (electron precipitation), where they deposit their energy by interacting with neutral atoms and cold charged particles. EMIC-driven electron precipitation still needs to be fully characterized and understood. In this work, we use data from the ELFIN CubeSats, which provide electron fluxes at high energy and pitch-angle (look direction) resolution at ~450 km of altitude. Our analysis reveals that precipitation is most efficient for ~MeV electrons and is accompanied by weaker low-energy precipitation down to ~100–200 keV. Given the ELFIN CubeSats spin, we can also study the distribution of the precipitating electrons along different look directions (pitch-angles). We find that the loss cone shape is well-reproduced by quasilinear predictions of EMIC-electron interactions at higher energies (≳250 keV), while quasilinear calculations underestimate the observed low-energy precipitation.

## 1 Introduction and Motivation

Electromagnetic Ion Cyclotron (EMIC) waves can cause depletions of outer radiation belt electrons at ~MeV energies (e.g., Blum et al., 2020; Capannolo et al., 2018; Drozdov et al., 2019, 2020; Shprits et al., 2017; Xiang et al., 2017; Zhang et al., 2016a). These electrons are lost through precipitation into the Earth's atmosphere due to pitch-angle scattering (Blum et al., 2015; Hendry et al., 2016; Li et al., 2014; Woodger et al., 2018). Flux variations in the radiation belt environment and the atmospheric impact of particle precipitation are essential for predicting and mitigating space weather. Charged particles in the radiation belts can potentially damage satellite electronics or instrumentation leading to anomalies or failures (Baker et al., 2018; Choi et al., 2011; Lotoaniu et al., 2015; Love et al., 2000; Shen et al., 2021), whereas precipitating electrons enhance ionospheric conductance and indirectly facilitate ozone reduction (Khazanov et al., 2018, 2021; Meraner & Shmidt, 2018; Randall et al., 2007; Yu et al., 2018). Therefore, it is crucial to quantify the effects of EMIC waves and characterize the resulting electron precipitation. This work aims to analyze Low-Earth-Orbit (LEO) observations of EMIC-driven electron precipitation from the recently completed ELFIN CubeSats mission (Electron Losses and Fields InvestigatioN) and study its properties in location, energy and pitch-angle.

Previous studies have provided some helpful information on the typical EMIC-driven characteristics (see Section 2.3). However, our understanding of EMIC-driven electron precipitation is still incomplete and limited mainly by the available detectors onboard satellites or balloons. While the widely used NOAA POES (Polar Orbiting Environmental Satellites) and EUMETSAT MetOp LEO satellite constellation provide wide L-MLT coverage and electron flux (>30 keV) along two looking directions (Carson et al., 2012; Shekhar et al., 2017, 2018), they are significantly affected by proton contamination (Capannolo et al., 2019a; Yando et al., 2011) when proton precipitation is strong, typically the case during EMIC wave activity (Cao et al., 2016; Miyoshi et al., 2008; Summers et al., 2017). Therefore, although proton contamination removal techniques exist (Hendry et al., 2017; Peck et al., 2015; Pettit et al., 2021), it is challenging to obtain reliable flux measurements of ≲700 keV electrons from POES data. Additionally, the energy channels are integral, preventing us from identifying the minimum energy of the precipitating electrons, even for uncontaminated events with reliable ≲700 keV flux. Improved observations were possible thanks to the FIREBIRD-II (Focused Investigations of Relativistic Electron Burst Intensity, Range and Dynamics; Johnson et al., 2020) CubeSats, which provided electron flux in differential energy channels at much higher energy and time resolution than POES. FIREBIRD's advantage is to directly probe 100s keV electrons without any proton contamination, revealing that EMIC waves can scatter ~200–300 keV electrons into the atmosphere together with ~MeV electrons (Capannolo et al., 2019b, 2021). However, due to its unstable and limited look direction, identifying FIREBIRD's pointing direction and quantifying the precipitation efficiency are challenging. Therefore, the number of reliable precipitation events observed by FIREBIRD is limited.

These instrumentation limitations can be overcome by the ELFIN CubeSats, specifically designed to investigate EMIC-driven precipitation at several energy channels and pitch-angles (both inside and outside the local loss cone). Angelopoulos et al. (2022) studied ~50 EMIC-driven events with a precipitation peak at >0.5 MeV from ELFIN. Their results are consistent with quasilinear theory: the energy of the precipitation peak is a proxy for the resonance energy at the frequency of peak wave power and the minimum resonance energy calculated from theory corresponds to the energy of the half-peak of the precipitation observed by ELFIN. In this letter, we extend this study to a larger dataset of events, independent of the spectral shape and concurrent with proton precipitation, a known signature of EMIC waves. We quantify the efficiency of electron precipitation, its energy spectrum, and the pitch-angle profile within the loss cone, and we compare the results with quasilinear simulations.

## 2 Data Description and Selection of MeV Precipitation Events

### 2.1 ELFIN

The twin ELFIN CubeSats were launched in September 2018 into a polar LEO (93° inclination, 450 km altitude) and have provided data from July 2019 to September 2022. The electron detector measures energies over 50 keV–8 MeV (16 logarithmically spaced channels). The CubeSats spun every 3 seconds and provided flux measurements of the entire 180° pitch-angle distribution (22.5° resolution), thus resolving electrons that are perpendicular (quasi-trapped: primarily within the drift loss cone at ELFIN's altitude), parallel (precipitating) and anti-parallel (mostly backscattered by the atmosphere) with respect to the magnetic field direction (Angelopoulos et al., 2020). Perpendicular electrons have pitch-angles $\alpha$ between the loss cone ($\alpha_{LC}$) and the anti-loss cone ($\alpha_{ALC}$). In the Northern Hemisphere, precipitating (backscattered) electrons have pitch-angles $\alpha < \alpha_{LC}$ ($\alpha > \alpha_{ALC}$). The opposite is true in the Southern Hemisphere. Electrons within the anti-loss cone have been backscattered by the atmosphere (e.g., Marshall & Bortnik, 2018; Selesnick et al., 2004) and will precipitate at the conjugate location in the opposite hemisphere, thus must be subtracted from the precipitating electron flux if we want to isolate the effects of wave-driven pitch-angle scattering and discard atmospheric scattering. We estimate the net precipitation fluxes as $J_{netprec} = J_{prec} - J_{back}$ (Mourenas et al., 2021, 2022). Additionally, to remove noise (low counts) from the data, we only use data for which the maximum percentage error is 50%.

## 2.2 POES/MetOp

The POES/MetOp (hereafter simply POES) constellation of LEO (~800–850 km altitude) spacecraft covers various MLT sectors and L shells (Evans & Greer, 2004; Sandanger et al., 2015). They are equipped with proton and electron detectors (MEPED, Medium Energy Proton and Electron Detector), thus suitable for studying EMIC-driven precipitation (e.g., Carson et al., 2012; Clilverd et al., 2015; Miyoshi et al., 2008). The 0° and 90° telescopes probe precipitating (deep inside the loss cone) and quasi-trapped (just outside the loss cone) particles at mid-to-high latitudes (Nesse Tyssoy et al., 2016, 2019), respectively. Precipitation is strong if the 0° flux is comparable to the 90° flux, indicating that a comparable fraction of trapped particles is precipitating. For our analysis, we primarily consider the proton differential energy channels P1 (30–80 keV), P2 (80–240 keV) and P3 (240–800 keV). We also use the relativistic electron channel E4 (>700 keV), obtained from the proton channel P6 as described in Green (2013) and Yando et al. (2011). The omnidirectional P6 proton channel (>16 MeV) provides reliable >3 MeV electron count rates when no protons are detected at higher proton energy (>35, >70, >140 MeV) (Evans et al., 2008; Sandanger et al., 2009).

## 2.3 Typical Signatures of EMIC-Driven Precipitation

Our goal is to identify electron precipitation events from ELFIN data potentially driven by EMIC waves. Several studies have indicated the typical signatures of EMIC-driven depletion and precipitation (Bruno et al., 2022; Capannolo et al., 2019a, 2021, 2022; Carson et al., 2012; Gasque et al., 2021; Hendry et al., 2016, 2017; Lyu et al., 2022; Miyoshi et al., 2008; Shekhar et al., 2018; Usanova et al., 2014; Zhang et al., 2016a). EMIC waves typically cause depletions and subsequent precipitation isolated in L shell, mainly observed from post-noon to post-midnight, more intense for high-energy electrons but also present for ~100s keV, coincident with isolated proton precipitation/aurora, and with a narrowing equatorial pitch-angle distribution of radiation belt electrons around 90°. Thus, we search for ELFIN observations of strong precipitation in the ~MeV range, occurring within the outer radiation belt and isolated in space. To strengthen the hypothesis that EMIC waves drive such precipitation, we also look for proton precipitation observed by POES (within ~1.5 L and ~3 MLT difference) where ELFIN observed precipitation. We discard any proton precipitation due to the curvature of magnetic field lines, which marks the proton isotropic boundary (PIB), identifiable given its L-shell and energy dependency (Delcourt et al., 1996; Dubyagin et al., 2018; Ganushkina et al., 2005; Gilson et al., 2012; Ebihara et al., 2011; Yu et al., 2020). All the events considered in this study occurred coincidentally with isolated proton precipitation observed at L-shells lower than the PIB.

We compare ELFIN observations with POES electron precipitation data (E4 and omnidirectional P6), though electron precipitation is not a requirement for event selection due to the high noise level in POES data which may conceal weaker electron precipitation (Nesse Tyssoy et al., 2016). We required that the precipitating-to-perpendicular ratio (prec-to-perp) observed by ELFIN during precipitation increases as a function of energy below ~1 MeV, as pitch-angle scattering is more efficient at higher energies (e.g., Capannolo et al., 2019a; Li et al., 2007; Summers & Thorne, 2003). A ratio that decreases to reach a minimum at 100s keV, followed by an increase could potentially indicate the coexistence of precipitation driven by chorus/hiss and EMIC waves (Angelopoulos et al., 2022). Typically, chorus and hiss are more efficient at driving electron precipitation at lower energies than that scattered by EMIC waves (Ma et al., 2016; Mourenas et al., 2021; Shprits & Ni, 2009; Summers et al., 2007). By following these criteria, we visually identified 144 events potentially driven by EMIC waves from all the available ELFIN data (July 2019 to September 2022). Each event showed precipitation of high-energy electrons, a prec-to-perp ratio increasing with energy, and 10s–100s keV proton precipitation nearby. We acknowledge that this is not a complete set of EMIC-driven precipitation observed during ELFIN's lifetime due to orbits precession of POES and ELFIN. Additionally, clear proton precipitation is observed by POES only within the PIB leading to some potential selection bias. Without other in-situ wave observations, it is impossible to completely rule out contributions from other wave types near the conjunction region. Nevertheless, our collection of events enables us to draw some conclusions, described as follows.

Figure 1 shows an example of a typical EMIC-driven precipitation event, with electron observations from ELFIN in panels a–g. Panels a–d illustrate the pitch-angle distribution in different energy ranges observed by ELFIN-A. ELFIN-A was in the Southern hemisphere: electrons with $\alpha < \alpha_{ALC}$ (dashed horizontal line) were backscattered by the atmosphere, those with $\alpha > \alpha_{LC}$ (solid horizontal line) precipitated in the Southern hemisphere. The perpendicular (quasi-trapped) population (panel e), $J_{perp}$, is obtained by averaging the flux at pitch-angles between $\alpha_{ALC}$ and $\alpha_{LC}$, and the precipitating population, $J_{netprec}$, is calculated by averaging the flux at pitch-angles $\alpha > \alpha_{LC}$ and subtracting the backscattered flux (as described in Section 2.1; $J_{back}$). Panel g shows the prec-to-perp ratio ($R = J_{netprec}/J_{perp}$) as a function of energy. Precipitation is marked by a horizontal gray bar. ELFIN-A observed clear isolated intense ~MeV electron precipitation accompanied by weaker low-energy precipitation down to ~150 keV (R~0.1), as also shown in Figure S1 in the Supporting Information (SI). MetOp-02 crossed a similar L-MLT region (~5mins later and ~2h of difference in MLT) and detected isolated precipitation of 10s–100s keV protons (h, gray bar), clearly distinct from the PIB (over ~02:04:35–02:04:45 UT) as well as electron precipitation in the >700 keV channel (i) and >3 MeV omnidirectional detector (j). For completeness, here we also show the E3 (>300 keV, green, panel i) electron channel (not affected by proton contamination in this case) to highlight that POES also detected precipitation at energies <700 keV (as observed by ELFIN-A). However, the integral POES make it challenging to identify the minimum energy of the precipitating electrons from POES data alone. Ground-based signatures of EMIC waves were also observed in a similar L-MLT region and UT by the Island Lake (ISLL, panel k) station in the Canadian Array for Realtime Investigations of Magnetic Activity (CARISMA, Mann et al., 2008) and other stations (e.g., ATH, Figure S2 in the SI).

## 3 Statistical Properties: Location, Minimum Precipitation Energy and Energy Trend

Figure 2 shows the statistical properties of the 144 EMIC-driven events observed by ELFIN. Panel a shows the distribution of the events (black circles) in L-MLT (using the T89 magnetic field model; Tsyganenko et al., 1989) with errorbars indicating the L-shell extent ($\Delta$L) of the precipitation. Errorbars for events with $\Delta$L ≤ 0.08 (as large as the dot symbols) are not shown and those with $\Delta$L > 0.5 are grayed out as they might indicate inaccurate magnetic field mapping rather than actual widespread precipitation. Most of the precipitation was observed over 5–8 L. The location of events outside this range, particularly those at high L shells, might be inaccurate due to the uncertainty in the magnetic field mapping, which could be less precise on the nightside. Figure S3 in the SI shows the L-MLT scatter plot with the IGRF magnetic field model and the magnetic latitude (MLAT) for each event. Precipitation was mainly observed over the dusk-to-midnight sector (15–24 MLT), but sometimes occurred to post-midnight and rarely towards noon. This distribution agrees with previous studies (Capannolo et al., 2019a, 2021; Gasque et al., 2021; Hendry et al., 2016; Shekhar et al., 2017; Yahnin et al., 2016, 2017): EMIC-driven precipitation often occurs where the conditions for efficient pitch-angle scattering are met (i.e., high plasma density and low magnetic field; Jordanova et al., 2008; Meredith et al., 2003; Silin et al., 2011; Summers & Thorne, 2003; Qin et al., 2020). We estimated the radial extent ($\Delta$L) of the precipitation by identifying regions of enhancement of $J_{netprec}$ ($\gtrsim 10^3$ electrons/s/sr/cm$^2$/MeV) and a high ratio ($\gtrsim 0.1$). These thresholds are subjective, thus the $\Delta$L estimate could vary. Here, we attempt to provide an upper boundary of the radial extent of precipitation given the 3s ELFIN time resolution. The observed precipitation often occurs on small radial scales (average $\Delta$L~0.3 corresponding to ~5s of precipitation; $\Delta$L histograms are in Figure S4), in agreement with previous studies (e.g., Capannolo et al., 2021; Lessard et al., 2019; Woodger et al., 2018) and likely due to the radially localized nature of EMIC wave excitation (e.g., Blum et al., 2017). These findings remain consistent also when using a more realistic magnetic field model (T05; Tsyganenko and Sitnov, 2005; not shown). To compare with previously studied EMIC-driven events, we also show the dataset of FIREBIRD (triangles) & POES (diamonds) conjunctions during EMIC waves from Capannolo et al. (2021). Colors indicate if precipitation occurred only at >700 keV (orange) or also at <700 keV (blue). Although this dataset is limited and no other events were identified at other MLTs during FIREBIRD & POES conjunctions, these observations agree with the duskside precipitation observed by ELFIN.

Regarding the minimum energy ($E_{min}$) of precipitating electrons observed by ELFIN, it appears that low-energy precipitation events (with ratio $\gtrsim 0.15$) typically occur over 15–24 MLT and sometimes over 0–4 MLT, as in Angelopolous et al. (2022). However, it is difficult to identify an isolated L-MLT region where low-energy precipitation is systematically observed. Figure 2b illustrates the percentage of events with $E_{min}$ lower than a certain threshold (colors) as a function of efficiency (R, y-axis). This figure shows that, for strong precipitation (high R), only a small percentage of events shows precipitation at low energy. For example, for a fixed energy threshold (green; $E_{min} \leq 500$ keV), the majority of events (55%) have $E_{min} \leq 500$ keV for ratio R~0.10, but only 20% of them have $E_{min} \leq 500$ keV for R~0.3. In other words, low-energy precipitation can occur, but it is much weaker than that at high energies. This is also clearly illustrated in Figure 2c, which displays R in each energy bin for the lower, middle (median) and upper quartiles of the dataset. No observations are shown for $\lesssim 90$ keV and $\gtrsim 2.5$ MeV due to the poor statistics in these bins, as indicated by the counts in the inset plot. This figure shows that the ratio monotonically increases with energy, in agreement with EMIC waves being more efficient at precipitating high-energy electrons. For the median value, R ~ 0.5 at ~1 MeV, but decreases to ~0.1 at ~240 keV. Observations of low-energy electron precipitation and high-energy electron precipitation in association with EMIC waves were also reported before (Capannolo et al., 2019b, 2021; Hendry et al., 2017, 2019), though our study further quantifies the energy dependence of the electron precipitation efficiency.

## 4 Average Pitch-angle Distribution in the Loss Cone

One of the key advantages of using ELFIN data is the unprecedented opportunity of studying the electron pitch-angle distribution at low altitudes, where the loss cone is sufficiently large to be resolved. In this section, we analyze the average electron flux as a function of pitch-angle during the selected precipitation events. Figure 3a shows the average observed flux J at each local pitch-angle, normalized by the flux at 90° ($J_{90°}$), as a function of pitch-angle (x-axis) and energy (colors). The local average loss cone angle ($\alpha_{LC}$) is shown with a vertical dashed line, delimiting the loss cone ($\alpha < \alpha_{LC}$). We obtained the average normalized flux at a fixed energy by first calculating the net flux at each pitch-angle by removing the backscattered electron flux as $J_{net}(\alpha) = J(\alpha) - J(\alpha_C)$, where $\alpha_C$ is the complementary pitch-angle in the anti-loss cone. The anti-loss cone fluxes are overall lower than the loss cone ones, indicating that the majority of precipitating electrons are not efficiently backscattered by the atmosphere. Note that backscattered electrons might have different energies and pitch-angle properties compared to precipitating electrons (Marshall & Bortnik, 2018; Selesnick et al., 2004); however, as a first approximation, we subtract the backscattered flux from the precipitating flux at the same energy and pitch-angle channel. Then, we calculated the normalized flux for each event, binned the dataset into 12° pitch-angle bins and averaged the values in each bin. Due to limited statistics (see Figure S5), we neglect the pitch-angle distribution at energies >2.5 MeV and pitch-angles close to 0°. Figure 3a shows that, on average, the loss cone fills up with increasing energy: it becomes half-full (R~0.5) for energies ≳ 1 MeV (similar to Figure 2c), though it is partly filled at low energies as well (~few 100s keV). Each point in Figure 3a is accompanied by a standard deviation shown in Figure S5, which indicates the data spread at each energy and pitch-angle. Thus, the negative gradient of the normalized flux at certain energies might not be realistic and should be analyzed more carefully in each event. This is the first time that such high energy and pitch-angle resolution has been used to resolve the loss cone during EMIC-driven precipitation.

The observed pitch-angle distribution is supported by simulations of EMIC-driven precipitation from quasilinear theory (Figure 3b). The simulations used statistical observations of EMIC waves at 6.5 L with wave amplitude $B_w > 1$ nT (from Zhang et al., 2016b) to calculate the expected pitch-angle distribution of electrons scattered by EMIC waves. More details are provided in Figure S6. The overall trend of a loss cone filling up with increasing energy is reproduced, especially ≳250 keV. However, quasilinear theory partly underestimates the contribution of low-energy electrons, potentially due to other effects (nonresonant interactions, nonlinear effects, etc.) not included in the quasilinear assumptions. The EMIC wave statistical properties, which affect the pitch-angle diffusion coefficient, may also be different from those driving the precipitation observed by ELFIN. Angelopoulos et al. (2022) suggested that precipitation at ~200–300 keV could be due to quasilinear scattering by moderately intense EMIC waves close to the proton gyrofrequency, yet low enough to evade hot plasma effects (Chen et al., 2011; Ross et al., 2021). Recent theories and simulations (An et al., 2022; Chen et al., 2016) also suggest that nonresonant interactions might explain the low energy EMIC-driven precipitation. Although we carefully ruled out events with a ratio displaying a minimum, other waves could still play some role in scattering low-energy electrons. However, without multiple in-situ wave observations and careful one-to-one conjunction analyses, it is challenging to exclude their contribution. Our analysis aims to provide the results of potentially EMIC-driven electron precipitation events carefully selected to discard doubtful cases, without explaining in detail why quasilinear simulations underestimate low-energy precipitation, which is beyond its scope.

**5 Summary and Conclusions**

We analyzed 144 electron precipitation events potentially driven by EMIC waves observed by the ELFIN CubeSats from July 2019 to September 2022. This analysis extends the study by Angelopoulos et al. (2022), which only selected events based on their spectral shape. In addition to requiring an increasing prec-to-perp ratio as a function of energy (to exclude possible chorus/hiss precipitation), we specifically selected events concurrent with proton precipitation observed by POES as a proxy for EMIC wave activity. While our results are similar to Angelopoulos et al. (2022), our study provides new insights into the pitch-angle shape of EMIC-driven precipitation inside the loss cone. Our key findings are:

1) EMIC-driven precipitation is observed on localized scales (average $\Delta L\sim0.3$) mainly over 5–8 L and 15–24 MLT, sometimes extending to post-midnight and dawn sectors;

2) Electron precipitation is stronger at energies $\gtrsim 1$ MeV (corresponding to a median prec-to-perp ratio $\gtrsim 0.5$) and extends to ~100–200 keV, albeit with less efficiency;

3) Low-energy precipitation (ratio$\gtrsim0.15$) is primarily observed over 15–24 MLT, sometimes extending up to 4 MLT;

4) The shape of the average flux as a function of pitch-angle shows the loss cone being gradually filled with increasing energy, in agreement with quasilinear calculations of the expected precipitation from EMIC waves. However, quasilinear calculations likely underestimate electron precipitation at $\lesssim250$ keV.

The localized nature of the EMIC-driven precipitation and the predominant dusk-to-midnight location of the events suggest that EMIC waves are most efficient at driving precipitation in this region, consistent with previous analyses using LEO satellites (FIREBIRD, POES and ELFIN; e.g., Angelopoulos et al., 2022; Capannolo et al., 2021; Hendry et al., 2017) and balloons (BARREL; Li et al., 2014; Woodger et al., 2018). This work also confirms that low-energy precipitation accompanies MeV precipitation during EMIC-wave activity but further quantifies the precipitation efficiency at any energy from ~100 keV up to ~2.5 MeV. For the first time, ELFIN enables us to measure the flux distribution in the loss cone with energy and pitch-angle: the loss cone is partly filled at low energies and becomes fuller at $\gtrsim1$ MeV. The modeled pitch-angle distribution due to strong EMIC waves from previous statistics overall agrees with the ELFIN average results; however, it partly underestimates the contribution of $\lesssim250$ keV electrons probably due to effects not included in quasilinear assumptions (i.e., nonresonant interactions), the possible contribution of other waves such as whistler-mode hiss or chorus, or potential different EMIC wave properties associated with the ELFIN dataset.

The present study provides a database of precipitation events likely driven by EMIC waves that can be used for future statistical or case studies (available at https://doi.org/10.5281/zenodo.7697272). These results provide unprecedented electron precipitation data at LEO with significantly improved energy and pitch-angle resolution, thus are particularly useful for radiation belt studies to quantify the effective wave-driven scattering. Additionally, our results are valuable for atmospheric studies to quantify the contribution of electron precipitation. In fact, models like BERI (Boulder Electron Radiation to Ionization; Xu et al., 2020) require information of the pitch-angle and energy electron distribution, which ELFIN can provide. We plan to use BERI to quantify the atmospheric ionization rates due to EMIC waves using ELFIN, overcoming the need to model the precipitating flux (Ma et al., 2022; Sanchez et al., 2022). Finally, this work joins previous analyses successfully conducted using CubeSat data (FIREBIRD: Breneman et al., 2017; Crew et al., 2016; Capannolo et al., 2019a, 2021; Duderstadt et al., 2021; Johnson et al., 2021; Shumko et al., 2018; AC6: Shumko et al., 2020a, 2020b; ELFIN: An et al., 2022; Artemyev et al., 2022; Gan et al., 2023; Grach et al., 2022; Mourenas et al., 2021, 2022; Shen et al., 2023; Zhang et al., 2022; see also Spence et al., 2022 for a review), demonstrating the remarkable assets these low-budget and educational missions bring to the field.

## Acknowledgments

Research at Boston University was supported by NASA grants 80NSSC20K0698, 80NSSC20K1270, 80NSSC21K1312, 80NSSC20K0196 and NSF grants AGS-2019950, AGS-2225445. AA and XJZ acknowledge the NASA 80NSSC23K0403 grant. The authors acknowledge all members of the ELFIN and POES/MetOp teams for providing data. The authors thank I.R. Mann, D.K. Milling and the rest of the CARISMA team for data. CARISMA is operated by the University of Alberta, funded by the Canadian Space Agency.

## Open Research

Data analysis was conducted using the SPEDAS library, publicly available at http://spedas.org/wiki/index.php?title=Downloads_and_Installation. ELFIN data are available at https://data.elfin.ucla.edu/ and processed using SPEDAS routines specifically written for processing ELFIN data by the ELFIN UCLA team. POES/MetOp data are available at: https://www.ncei.noaa.gov/data/poes-metop-space-environment-monitor/access/l1b/v01r00/. Data from the ground-based CARISMA station are available at http://www.carisma.ca/themis_carisma_cdf/2021/02/02/. Data for the Athabasca station in the Supporting Information is available at https://ergsc.isee.nagoya-u.ac.jp/data/ergsc/ground/geomag/isee/induction/ath/2021/02/. The dataset of the analyzed events is available at https://doi.org/10.5281/zenodo.7697272.

**Figure 1**. Overview of an example of precipitation observed by ELFIN and POES, during EMIC waves observed at ground. ELFIN observations: (a–d) pitch-angle distribution at different energies with local loss (anti-loss) cone marked by a solid (dashed) line, (e) quasi-trapped and (f) precipitating electrons (backscattered subtracted), and (g) prec-to-perp ratio. POES observations: (h) proton and (i) electron flux at different energies (colors) from the 90° (dotted) and 0° (solid) telescopes, and (j) count rate from the OMNI proton channel, strongly contaminated by >3 MeV electrons. Precipitation is indicated by a gray bar in panels (g), (h)–(j). (k) An example of EMIC waves observed at the Island Lake station.

**Figure 2**. (a) L-MLT distribution of the electron precipitation events: black dots, triangles and diamonds are from ELFIN, FIREBIRD and POES observations, respectively, with errorbars indicating the L-shell extent ($\Delta L$) of the precipitation. $\Delta L \leq 0.08$ are not shown and $\Delta L > 0.5$ are marked in gray to indicate potentially inaccurate field mapping. FIREBIRD and POES events are from Capannolo et al. (2021) and colored in blue or orange depending on the minimum energy of the observed electron precipitation. (b) Percentage of events with $E_{min} \leq 300$ keV (blue), 500 keV (green), 700 keV (orange) for a given prec-to-perp ratio (R, y-axis). For example, 20% of the events revealed precipitation at $\leq 300$ keV with R~0.15. (c) R as a function of energy for the lower, middle and upper quartiles. The inset plot shows the number of points in each energy bin and indicates that energies with <45 points have been discarded.

**Figure 3**. Normalized flux ($J/J_{90°}$) as a function of local pitch-angle and energy (colors): (a) average ELFIN observations (12° bins), (b) quasilinear results of expected EMIC-driven precipitation (EMIC waves at 6.5 L and amplitudes >1 nT; Zhang et al., 2016b). The bounce loss cone angles in panels (a) and (b) are from the IGRF and dipole magnetic fields, respectively.

# ELFIN-A

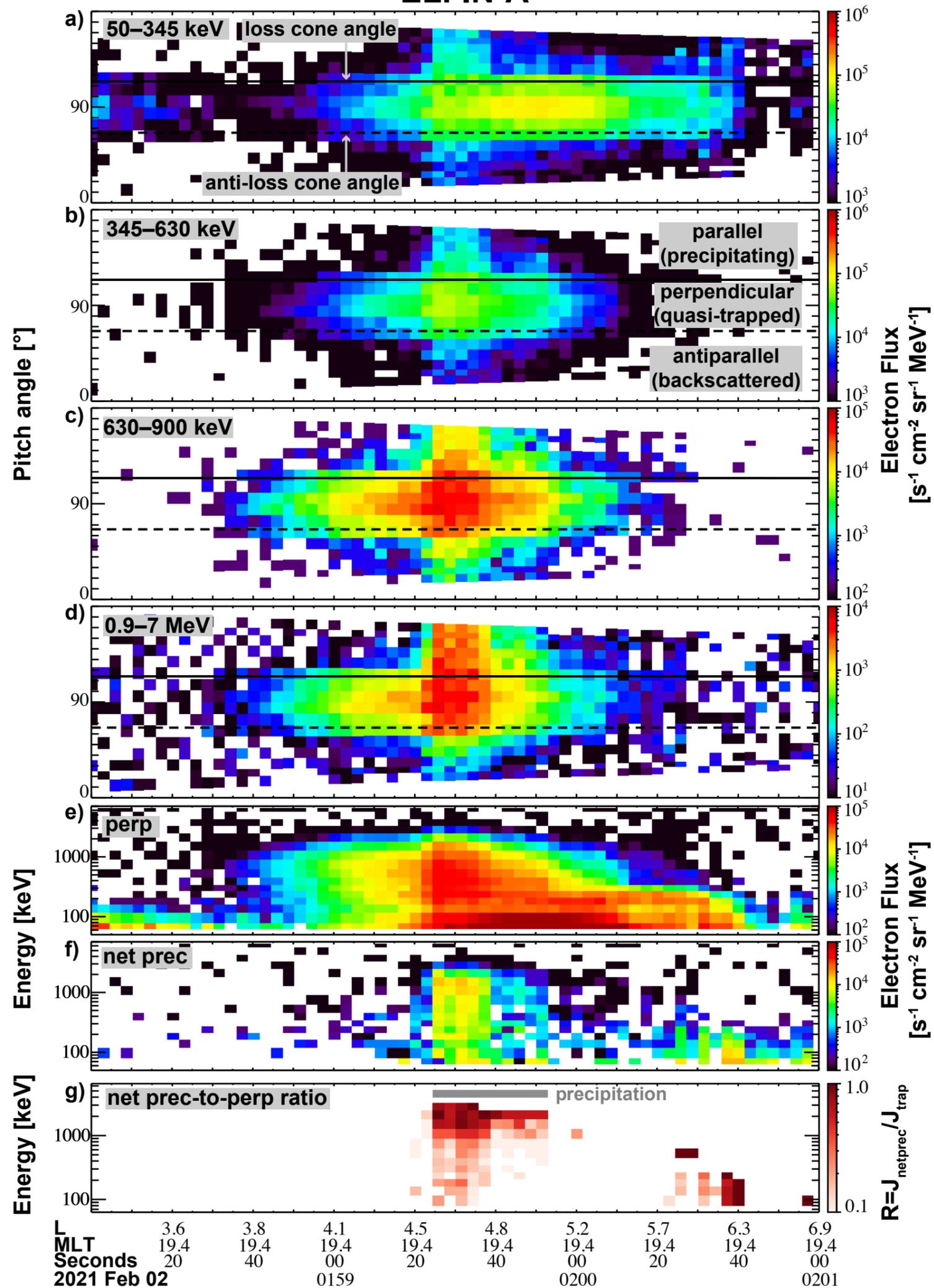

a) 50–345 keV  loss cone angle  anti-loss cone angle

b) 345–630 keV  parallel (precipitating)  perpendicular (quasi-trapped)  antiparallel (backscattered)

c) 630–900 keV

d) 0.9–7 MeV

e) perp

f) net prec

g) net prec-to-perp ratio  precipitation

Pitch angle [°]

Energy [keV]

Electron Flux [s$^{-1}$ cm$^{-2}$ sr$^{-1}$ MeV$^{-1}$]

Electron Flux [s$^{-1}$ cm$^{-2}$ sr$^{-1}$ MeV$^{-1}$]

$R = J_{netprec}/J_{trap}$

L        3.6    3.8    4.1    4.5    4.8    5.2    5.7    6.3    6.9
MLT    19.4  19.4  19.4  19.4  19.4  19.4  19.4  19.4  19.4
Seconds 20  40    0159  20    40   0200  20    40   0201
2021 Feb 02

# MetOp-02

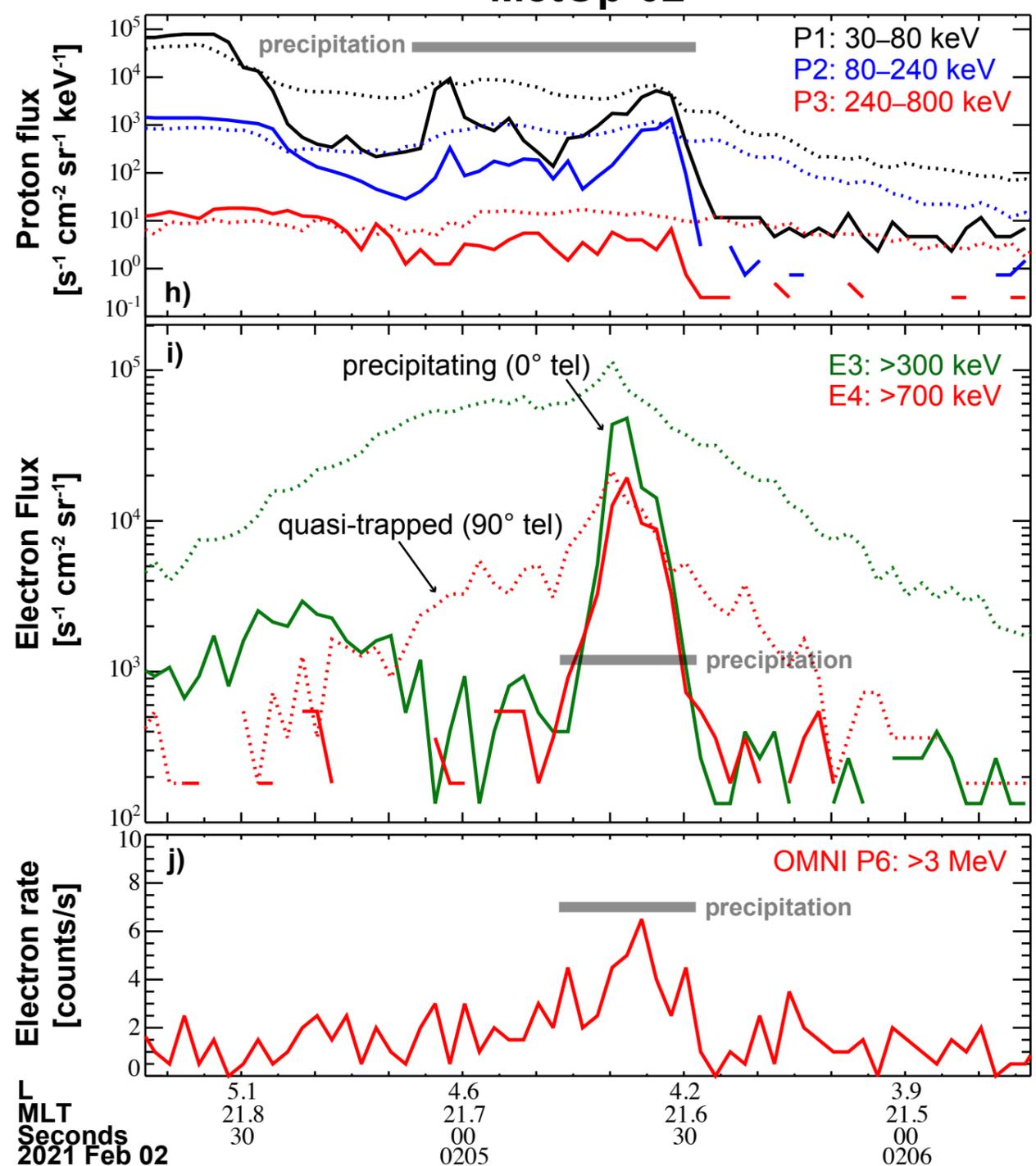

h)  precipitation  P1: 30–80 keV  P2: 80–240 keV  P3: 240–800 keV

Proton flux [s$^{-1}$ cm$^{-2}$ sr$^{-1}$ keV$^{-1}$]

i)  precipitating (0° tel)  quasi-trapped (90° tel)  E3: >300 keV  E4: >700 keV  precipitation

Electron Flux [s$^{-1}$ cm$^{-2}$ sr$^{-1}$]

j)  OMNI P6: >3 MeV  precipitation

Electron rate [counts/s]

L        5.1    4.6    4.2    3.9
MLT    21.8  21.7  21.6  21.5
Seconds 30    0205  00    30   0206
2021 Feb 02

# CARISMA/ISLL

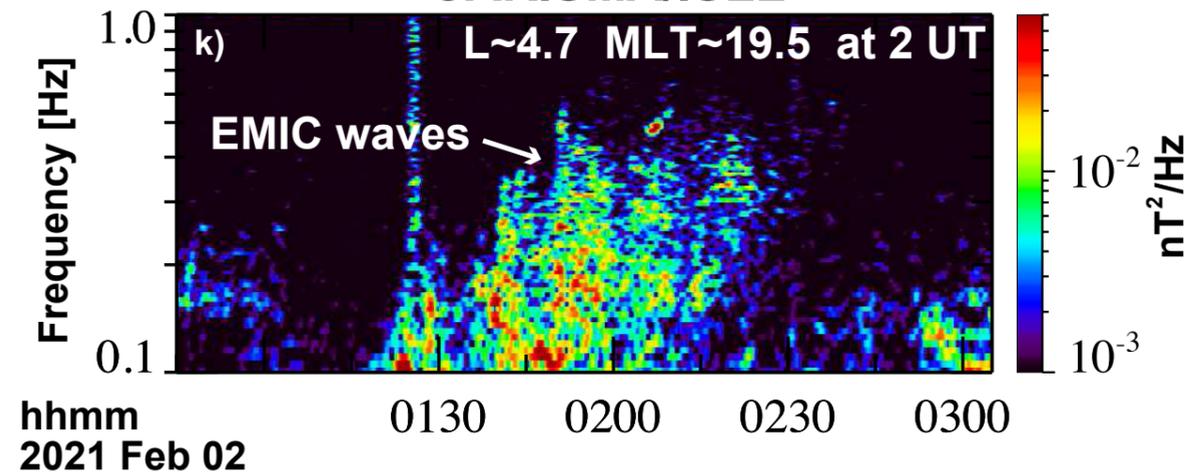

k) L~4.7  MLT~19.5 at 2 UT

EMIC waves

Frequency [Hz]

nT$^2$/Hz

hhmm  0130    0200    0230    0300
2021 Feb 02

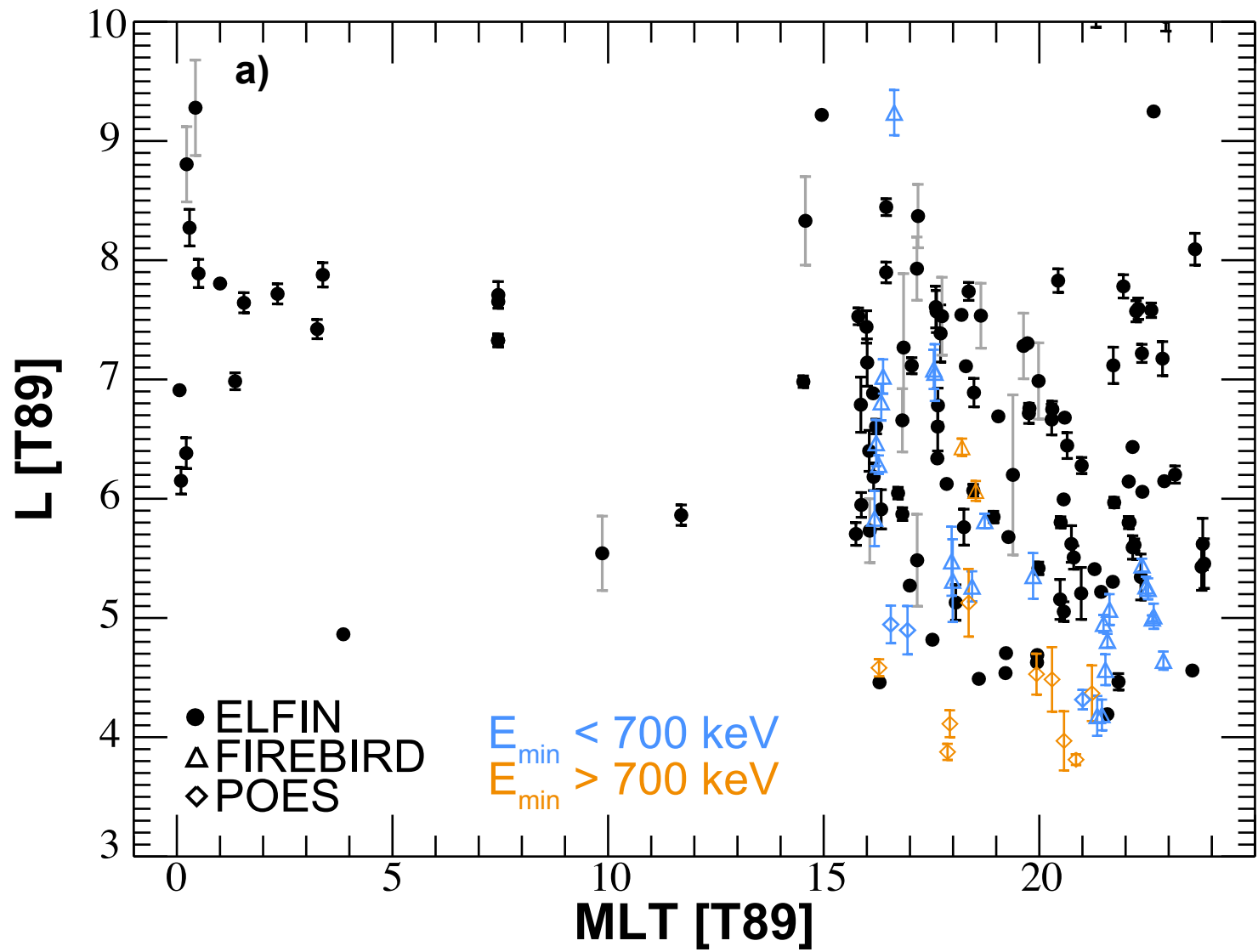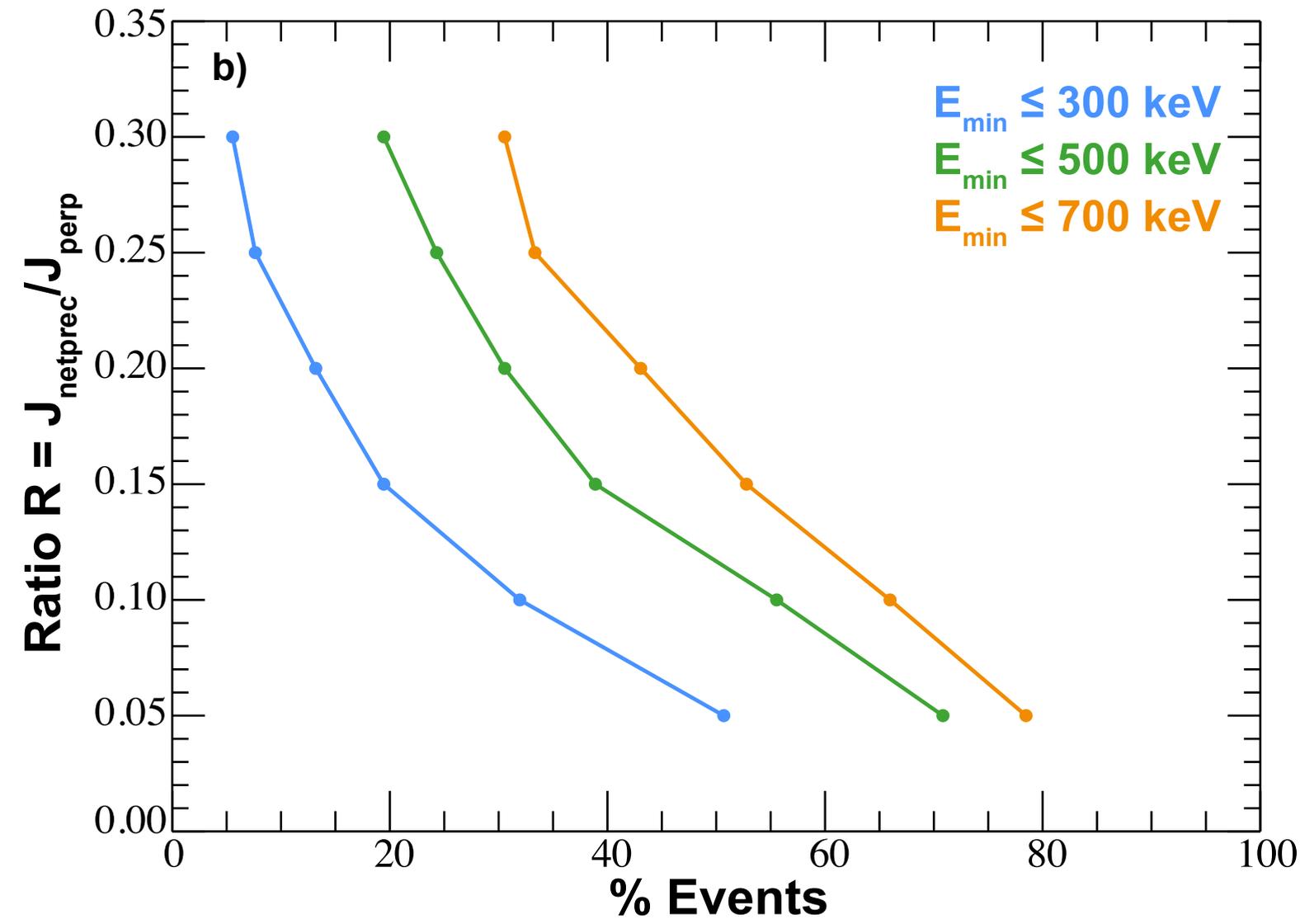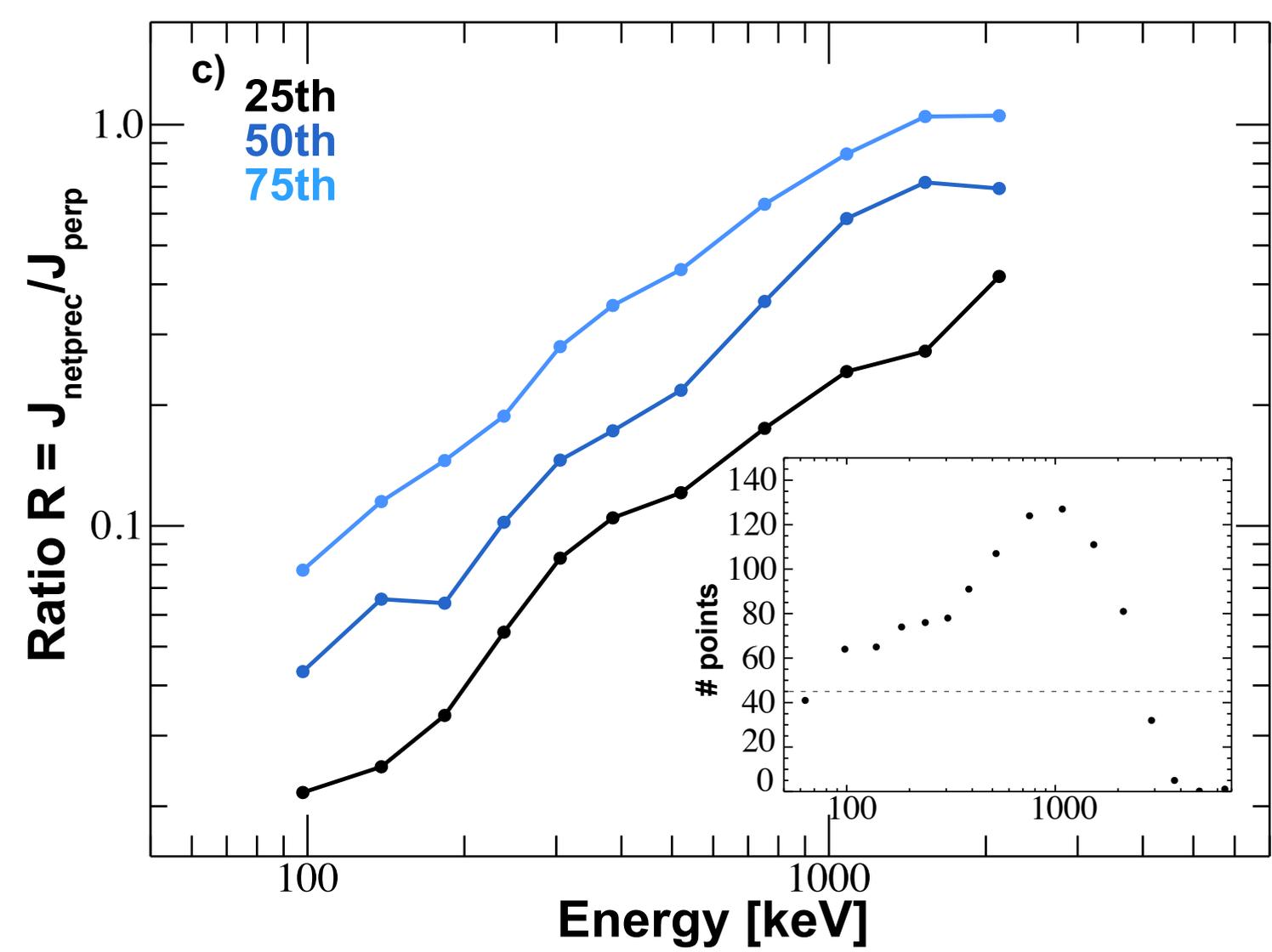

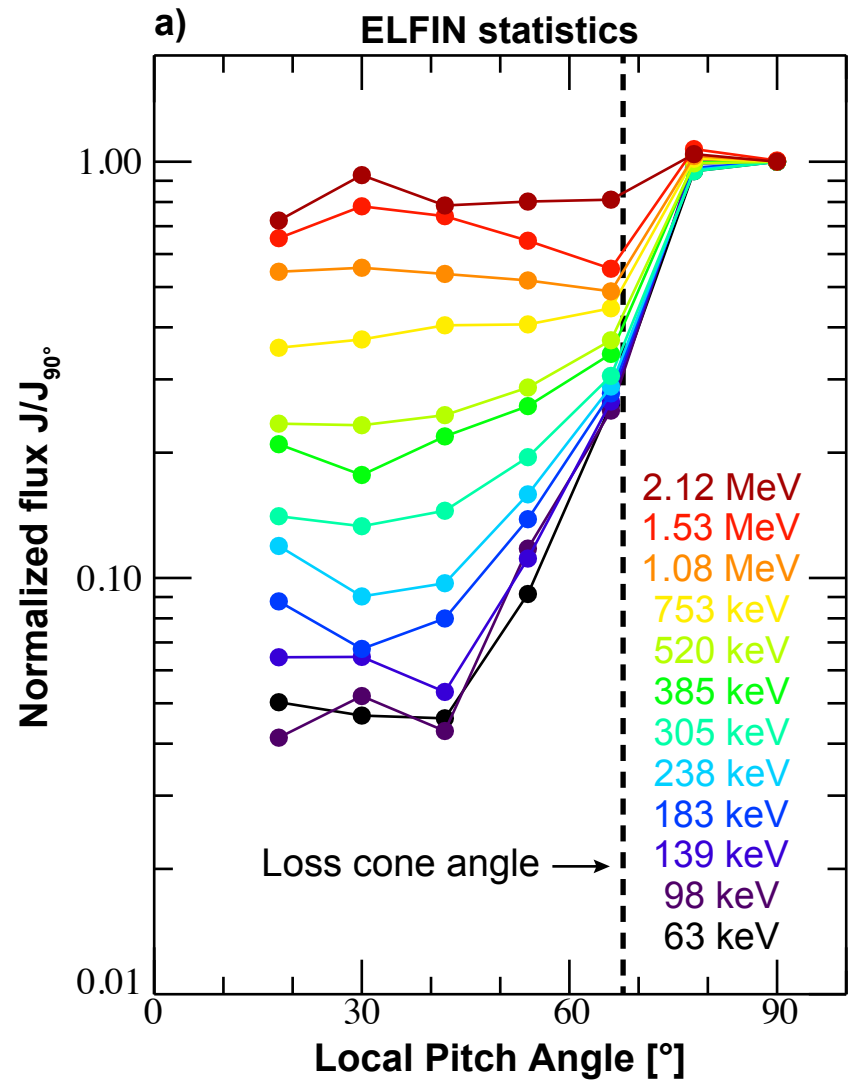

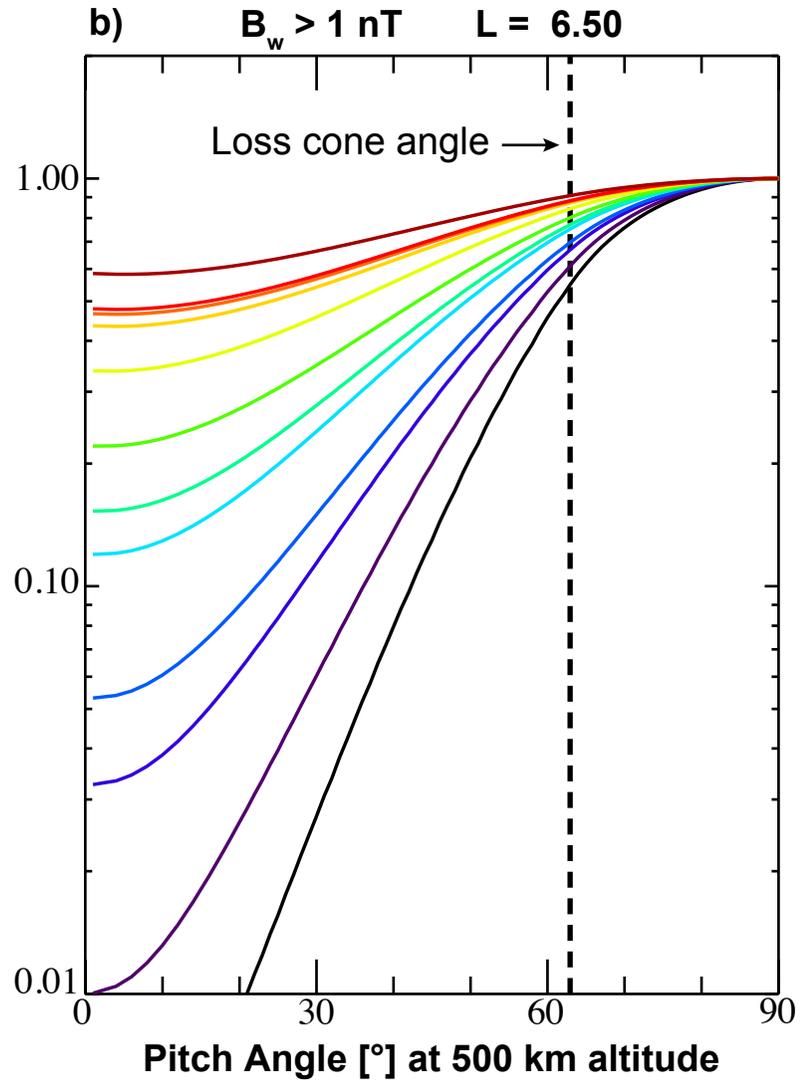